\documentclass[aps,pre,amssymb,amsmath,preprint]{revtex4}

\usepackage{upgreek}
\usepackage{rotating}
\usepackage{marvosym}
\usepackage{booktabs}
\usepackage{multirow}

\newcommand{\derpar}[2]{\dfrac{\partial{#1}}{\partial{#2}}}
\newcommand{\dsfrac}[2]{{\displaystyle\frac{#1}{#2}}}
\newcommand{\Int }{\displaystyle\int}

\newcommand{\mib}{\boldsymbol}
\newcommand{\fb}{{\mib f}}
\newcommand{\kappab   }{{\mib\kappa}}
\newcommand{\pb}{{\mib p}}
\newcommand{\ub}{{\mib u}}
\newcommand{\vb}{{\mib v}}

\newcommand{\HC}{{\mathcal{H}}}
\newcommand{\OC}{{\mathcal{O}}}

\newcommand{\av}[1]{\left\langle #1 \right\rangle}

\newcommand{\Gav}{\mathcal{G}}
\newcommand{\Ghav}{\widehat{\Gav}}

\newcommand{\Qav}{\mathcal{Q}}

\begin{document}

\preprint{}

\title{Direct-Numerical-Simulation-based measurement \\ of the mean impulse response \\
       of homogeneous isotropic turbulence}

\author{Marco Carini}
\email{carini@aero.polimi.it}
\author{Maurizio Quadrio}
\email{maurizio.quadrio@polimi.it}
\affiliation{Dipartimento di Ingegneria Aerospaziale
Politecnico di Milano, Campus Bovisa, Via La Masa 34, 20156 Milano, Italy}

\begin{abstract}
A technique for measuring the mean impulse response function of stationary homogeneous isotropic turbulence is proposed. Such measurement is carried out here on the basis of Direct Numerical Simulation (DNS). A zero-mean white-noise volume forcing is used to probe the turbulent flow, and the response function is obtained by accumulating the space-time correlation between the white forcing and the velocity field. This technique to measure the turbulent response in a DNS numerical experiment is a new research tool in that field of spectral closures where the linear response concept is invoked either by resorting to renormalized perturbations theories or by introducing the well-known Fluctuation-Dissipation Relation (FDR). Though the results obtained in the present work are limited to relatively low values of the Reynolds number, a preliminary analysis is possible. Both the characteristic form and the time scaling properties of the response function are investigated in the universal subrange of dissipative wavenumbers; a comparison with the response approximation given by the FDR is proposed through the independent DNS measurement of the correlation function. Very good agreement is found between the measured response and Kraichnan's description of random energy-range advection effects. 
\end{abstract}

%\pacs{XXX}

%\keywords{Homogeneous isotropic turbulence, statistical closure theories, impulse response measurement, fluctuation-response relation, DNS.}
\maketitle

%%%%%%%%%%%%%%%%%%%%%%%%%%%%%%%%%%%%%%%%%%%%%%%%%%%%%%%%%%%%%%%%%%%%%
\section{Introduction}

The concept of impulse response tensor of an isotropic turbulent flow lies at the heart of the Direct Interaction Approximation (DIA) theory, developed 50 years ago \cite{kraichnan-1959} by the great theoretical physicist Robert Kraichnan, to tackle the turbulence closure problem analytically. Since then, within the renormalized perturbations approach, several closure strategies have been proposed (a significant example is the Local Energy Transfer (LET) theory introduced by McComb \cite{mccomb-1974}), eventually adopting a Lagrangian viewpoint, as done by Kraichnan himself \cite{kraichnan-1964-a, kraichnan-1977} and others \cite{kaneda-1981, kida-goto-1997}. In all such theories, either Eulerian or Lagrangian, closure is achieved by means of a closed set of integro-differential equations, where the unknowns are the two-points, two-times velocity correlation tensor and the response tensor itself. An exception is LET, where the response tensor is replaced by a renormalized propagator tensor which connects the velocity correlations at different times, in close analogy with the well known fluctuation-dissipation relation of the classical statistical physics. Recently, McComb and Kiyani \cite{kiyani-mccomb-2004} have shown how a renormalized response tensor relating the two-point covariance at different times can be derived; the corresponding relationship reduces to a FDR form, still within the theoretical framework of second-order renormalized perturbations, by introducing the so called time ordering approach to reconcile the time-symmetry of the correlation with the causality of the response. 

During the last decades, several statistics of homogeneous isotropic turbulence (HIT), either computed with well resolved direct numerical simulations or obtained from experiments, have been compared to the corresponding theoretical predictions at increasing values of $Re_\lambda$ \cite{ishihara-gotoh-kaneda-2009}, in the statistically stationary as well as in the freely decaying regime. Encouraging results both for the LET theory and various Lagrangian closures have been reported \cite{mccomb-1990, mccomb-filipiak-shanmugasundaram-1992, kaneda-1993, kida-goto-1997}. Up to the present day, however, such a comparison for the impulse response function has never been addressed, owing to the lack of (experimental or numerical) information about it. Missing such a comparison is not a minor issue for Eulerian closure theories: as stressed in Ref. \cite{mccomb-1990}, the differences among the various theoretical approaches have their roots in the form of the response or propagator equation, whereas the covariance equation is most often treated in equivalent ways. Furthermore, if the response and the two-point covariance were available, the degree of approximation involved in using the FDR in the context of HIT could be straightforwardly evaluated, indirectly gathering information about the invariant probability distribution of the turbulent system \cite{biferale-etal-2002}.

In recent years Luchini {\em et al.} in Ref. \cite{luchini-quadrio-zuccher-2006b} have proposed an original method to carry out an Eulerian DNS-based measurement of the mean impulse response of a turbulent flow, and have described the response function of a fully developed turbulent channel flow to small-amplitude perturbations applied at the wall. That study was conceived in the framework of turbulence control (hence the emphasis on wall flows and wall forcing); due to lack of isotropy, the response tensor is quite complicated, and does not directly relate to the previous isotropic theories. However, the proposed measurement technique provides us with the required tools to obtain the impulse response tensor for HIT, where the response function shall be intended to describe the response of turbulence to volume forcing.  The present paper therefore aims at measuring the Eulerian HIT response, presenting preliminary results, obtained at low values of $Re_\lambda$, that will enable us to analyze the characteristic form and time scales of the response, and to compare them with theoretical predictions and assumptions.

The paper is organized as follows. In the next \S\ref{sec:measure}, the definition of the impulse response is briefly reviewed to introduce the measurement technique, that is numerically validated against the available analytical viscous solution, and to discuss accuracy issues. In \S\ref{sec:results} the actual response function is presented and analyzed with reference to the theoretical background of renormalized perturbations and FDR. Lastly, \S\ref{sec:concl} is devoted to a concluding discussion.

%%%%%%%%%%%%%%%%%%%%%%%%%%%%%%%%%%%%%%%%%%%%%%%%
\section{Measuring the response function by DNS}
\label{sec:measure}

\subsection{The definition of the impulse response function}
\label{sec:resp-def}

Following Ref. \cite{sagaut-cambon-2008}, the most general definition in wave vector space $\kappab$ of the instantaneous impulse response tensor of a turbulent velocity field $\ub(\kappab,t)$ to an external volume force $\fb(\kappab,t)$, is given by the following input-output relationship between infinitesimal perturbations, $\updelta$ (note the different notation from the Dirac delta function $\delta(\cdot)$):
\begin{equation}
 \updelta u_i(\kappab,t) = 
 \Int \Int_{-\infty}^t H_{in}(\kappab,\kappab',t,t') \updelta f_n(\kappab',t') dt'd\kappab'.
\label{eq:resp-def}
\end{equation}

It is important to underline that perturbations here assume a stochastic meaning, since they are superimposed to a particular random realization of $\ub$, which itself is solution of the fully non-linear Navier-Stokes Equations (NSE) in Fourier space. Therefore $H_{in}(\kappab,\kappab',t,t')$ possesses a random nature, and an integral formulation not only in time but also in wave-vector space is required. In fact the instantaneous response tensor plays the role of a tangent Green's function, related to a random and nonlinear state, and satisfies the instantaneous response equation:
\begin{widetext}
\begin{equation}
   \left(\derpar{}{t} + \nu \kappa^2 \right) H_{in}(\kappab,\kappab',t,t') 
   = 2 M_{ijm}(\kappab) \Int u_j(\pb,t) H_{mn}(\kappab-\pb,\kappab',t,t') d\pb +
   P_{in}(\kappab') \delta(\kappab-\kappab') \delta(t-t'),
\label{eq:ist-resp}
\end{equation}
\end{widetext}
which can be derived through a stochastic Green function formalism applied to the linearized form of Fourier transformed NSE. In Eq. (\ref{eq:ist-resp})  $M_{ijm}(\kappab)$ is the inertial transfer operator given by:
\begin{equation}
 M_{ijm}(\kappab) = -i/2(\kappa_mP_{ij}(\kappab)+\kappa_jP_{im}(\kappab)),
\end{equation}
and $P_{ij}(\kappab)$ is the projection tensor in wave-vector space, expressed as:
\begin{equation}
 P_{ij}(\kappab) = \delta_{ij} -\kappa^{-2}\kappa_i\kappa_j. 
\end{equation}

The locality of the response tensor in wave-vector space follows only after averaging:
\begin{equation}
 \av{H_{in}} = \HC_{in}(\kappab,t,t')\delta(\kappab-\kappab').
\label{eq:resp-loc}
\end{equation}

Lastly, exploiting statistical isotropy and stationarity results in scalar response functions, respectively $\Ghav$ and $\Gav$, defined as follows:
\begin{equation}
 \HC_{in}(\kappab,t,t') = P_{in}(\kappab)\Ghav(\kappa,t,t'),
\label{eq:resp-isodef}
\end{equation}
\begin{equation}
 \Gav(\kappa,\tau) = \Ghav(\kappa,t,t-\tau).
\end{equation}
The causality property holds for both the previous functions, hence :
\begin{equation}
 \Gav(\kappa,\tau) = 0 \quad \mbox{for } \tau < 0 \mbox{ and } \forall \kappa .
\end{equation}
This is obviously a consequence of the realizability of the dynamical system that is being described through its impulse response. As indicated by Kraichnan \cite{kraichnan-1959}, the scalar response is a real, unit-bounded function:
\begin{equation}
|\Gav(\kappa,\tau)| \leq \Gav(\kappa,0^{+}) = 1, \quad \forall \tau > 0 \mbox{ and } \forall \kappa .
\end{equation}

%----------------------------------------------
\subsection{The Direct Numerical Simulation}
\label{sec:dnscode}

\begin{figure}
\centering
\includegraphics[width=0.45\textwidth]{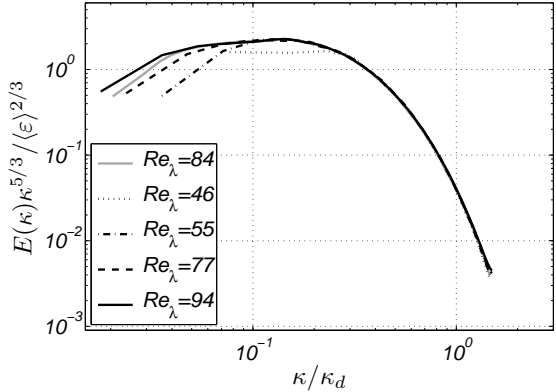}
\caption{Compensated energy spectrum for HIT: the function $E(\kappa)$ computed with the present DNS code at several values of $Re_\lambda$ is compared with results from Ref. \cite{lamorgese-caughey-pope-2005} at $Re_\lambda = 84$.}
\label{fig:spectra}
\end{figure}

The measurement of $\Gav$ described in this paper is carried out by means of a forced DNS of stationary HIT on a cubic domain, whose edge length $L$ is chosen to be $L = 2 \pi$ for convenience, so that the fundamental wave number is $\kappa_0 = 2 \pi / L = 1$ without loss of generality. A numerical code has been developed on purpose and equipped with parallel (shared-memory) computing capabilities. The code implements a classical Galerkin-Fourier scheme applied to the velocity-vorticity formulation of the incompressible Navier-Stokes equations. In the present context, this formulation presents interesting advantages in terms of memory requirements. Exact removal of the aliasing error is obtained with the 3/2 zero-padding rule; time integration is carried out by means of a third-order low-storage Runge-Kutta (Williamson) scheme; see Ref. \cite{canuto-etal-2006,canuto-etal-2007} for additional numerical details. The forcing scheme has been carefully implemented following the provisions stated in Ref. \cite{lamorgese-caughey-pope-2005}, from which the notation adopted below is borrowed. The Kolmogorov scale is indicated with $\eta$, with $\kappa_d = \eta^{-1}$, the instantaneous dissipation rate is $\varepsilon$, the forcing-containing shell is $\kappa_f$ and the mean energy injection rate is $P$, that equals $\av{\varepsilon}$ at statistical stationarity. Then the adopted feedback-acceleration forcing \cite{lamorgese-caughey-pope-2005} is formulated in wave number space as follows:
\begin{equation}\label{eq:forcing}
 \fb(\kappab,t) = \dsfrac{Ph(\kappab;\kappa_f)}{2 k_f(t)}\ub(\kappab,t),
\end{equation}
where $k_f(t)$ represents the kinetic energy of the modes within the forced shell $\kappa_f$ and $h(\kappab;\kappa_f)$ 
is the related indicator function:
\begin{equation}
 h(\kappab;\kappa_f) =
 \begin{cases}
    1, & \vert\kappab\vert \leq \kappa_f,\\[2mm]
    0, & \text{otherwise}.
 \end{cases}
\end{equation}
A standard resolution of $\kappa_{max} \eta = 1.5$ is adopted, where $\kappa_{max}$ indicates the maximum resolved wave-number in each direction of the Fourier space. The numerical code has been thouroughly verified by running conventional simulations of stationary HIT. The computed energy spectra at various $Re_\lambda$ compare very well to available results. A comparison of this kind is shown in Fig. \ref{fig:spectra}, that shows excellent agreement between our computed energy spectra and those pubblished in Ref. \cite{lamorgese-caughey-pope-2005}. The spectral code has been run on a machine equipped with 4 Opteron 2378 processors, where a case with $N=256$ has a memory requirement of 940MB and a typical execution time of 11 seconds for one Runge--Kutta time step.
%----------------------------------------------

\subsection{The response measurement technique}
\label{sec:technique}

In Ref. \cite{luchini-quadrio-zuccher-2006b} Luchini, Quadrio \& Zuccher propose an innovative method for measuring the linear impulse response of a turbulent velocity field, resorting to the statistical statement of the input-output relation for a linear system, i.e. {\em the input-output correlation}. This approach is primarily motivated by the problem of low signal-to-noise ratio (S/N) that one would face, should the response function be measured according to its definition. Indeed the linear response of a non-linear dynamical system is obtained by means of infinitesimal perturbations around an equilibrium state, which has a stochastic meaning in the description of turbulence. Therefore impulsive perturbations externally introduced into the turbulent field to measure its linear response must be extremely small compared to the natural turbulent fluctuations for Eq. (\ref{eq:resp-def}) to hold; as a consequence, their effect is buried into turbulent noise.

By definition the impulse response is the output of a linear system when either harmonic or impulsive signals are used as inputs. However, for a linearized turbulent system, the use of a proper statistical probe instead of a deterministic one will dramatically improve the computational efficiency of the overall measurement procedure. This is the case of using a white-noise process in input to the system. Indeed it is well known from filtering theory \cite{jazwinski-1970} that when a linear system is fed with white noise, the correlation between the input and the output is proportional to the impulse response of the system, owing to the delta-correlated property of the white-noise process. 
We employ an externally generated random volume forcing as the input; by computing its cross-correlation with the velocity field, the whole wave-number dependency of the response function is obtained at once. At the same time, forcing is uniformly distributed over time and space, thus leading to improved S/N and larger allowed amplitudes within the linearity constraint. Therefore this strategy performs much better than a deterministic forcing, be it either harmonic or impulsive, that would lead to computationally unaffordable simulations, as highlighted in Ref. \cite{luchini-quadrio-zuccher-2006b}.

Starting from Eq. (\ref{eq:resp-def}), the input-output correlation can be written as:
\begin{widetext}
\begin{equation}
   \av{\updelta u_i(\kappab,t) \updelta f_j(-\kappab,t-\tau)} 
   =\Int\Int_{-\infty}^{+\infty} \HC_{in}(\kappab,t-t') \delta(\kappab'-\kappab)
   \av{\updelta f_n(\kappab',t') \updelta f_j(-\kappab,t-\tau)} dt'd\kappab',
\label{eq:input-output-corr}
\end{equation}
\end{widetext}
where Eq. (\ref{eq:resp-loc}) has been used owing to the average operator, and the response causality property allows the extension towards $+\infty$ of the upper bound of time integral. Assuming $\updelta f_j(\kappab,t) = \epsilon w_j(\kappab,t)$, being $\epsilon\in\mathbb{R}^+$ a scale factor and $w_j(\kappab,t)$ an independently generated zero-mean white-noise field with identity covariance matrix:
\begin{equation}
 \av{\updelta f_n(\kappab,t')\updelta f_j(-\kappab,t-\tau)} = \epsilon^2\delta_{nj}\delta(t'-t+\tau),
 \label{eq:noise}
\end{equation}
the cross-correlation at the l.h.s. of Eq. (\ref{eq:input-output-corr}) will result in the properly scaled response tensor:
\begin{equation}
   \av{\updelta u_i(\kappab,t)\updelta f_j(-\kappab,t-\tau)} = \epsilon^2\HC_{ij}(\kappab,\tau).
\end{equation}

We shall denote by $\widetilde{\ub}(\kappab,t)$ the turbulent velocity field when volume forcing with white spectrum is applied. If the perturbation is small enough for linearity to hold, 
i.e. $\epsilon \ll 1$, it follows that:
\begin{equation}
 \widetilde{\ub}(\kappab,t) = \vb(\kappab,t) + \updelta\ub(\kappab,t),
\end{equation}
where $\vb(\kappab,t)$ indicates a different realization of the turbulent fluctuating field respect to the original field $\ub(\kappab,t)$, as a consequence of non-linearity and stochastic behavior of NSE. Then computing the correlation between $\widetilde{\ub}$ and $\updelta\fb$ results in:
\begin{widetext}
\begin{equation}
  \dsfrac{\av{\widetilde{u}_i(\kappab,t) \updelta f_j(-\kappab,t-\tau)}} {\epsilon^2}
  = \dsfrac{1}{\epsilon^2} \left[\av{v_i(\kappab,t) \updelta f_j(-\kappab,t-\tau)} 
  + \av{\updelta u_i(\kappab,t) \updelta f_j(-\kappab,t-\tau)} \right].
  \label{eq:input-output-corr-expl}
\end{equation}
\end{widetext}

Since the applied random perturbation on forcing is uncorrelated to turbulent fluctuations, the term $\av{v_i (\kappab,t)\updelta f_j(-\kappab,t-\tau)}$ will be averaged out in the previous equation, leading to:
\begin{equation}
 \dsfrac{\av{\widetilde{u}_i(\kappab,t) \updelta f_j(-\kappab,t-\tau)}}{\epsilon^2} =
 \HC_{ij}(\kappab,\tau),
 \label{eq:input-output-corr-last}
\end{equation}
where the input-output correlation law, Eq. (\ref{eq:input-output-corr}), has been used to handle the non-vanishing term (the second term) at r.h.s. of Eq. (\ref{eq:input-output-corr-expl}). In this way it is still possible to measure the turbulent response using the cross-correlation between the white-noise input and the whole turbulent velocity field. 

At this point it may be useful to note explicitely that no relation exists between the energy-driving forcing, Eq. (\ref{eq:forcing}), and the white-noise forcing applied for the response measurement, Eq. (\ref{eq:noise}). The former obviuosly represent a mere artificial, but unavoidable, mechanism to drive the flow and maintain statistical stationarity, via supply of energy at large scales. The white noise, on the other hand, is an external field of volume force that enters the stationary turbulent system, which includes the energy-driving forcing. The $j$-th component of the white noise field, $w_j(\kappab,t)$ is defined as:
\begin{equation}
 w_j(\kappab,t) = \exp(i 2 \pi \phi),
\end{equation}
where $\phi$ is the output of a random number generator with an uniform probability distribution in the interval $[0,1]$ \footnote{Note that a white noise field with identity covariance matrix is used in the definition of $\updelta f_j(\kappab,t)$, Eq. (\ref{eq:noise}): the identity covariance scaling factor for $w_j(\kappab,t)$ could be easily embedded in the definition of $\epsilon$ without losing of generality.}. Hence the white noise is independent from both the turbulent fluctuations and the energy-driving forcing applied to them through the feedback formula (\ref{eq:forcing}). A feedback forcing loop should be considered when looking at the linear response for wave numbers contained in the forced shell, but, as already discussed, these are not of physical interest.

In the HIT case Eq. (\ref{eq:resp-isodef}) provides us with a convenient way of accumulating just the simple scalar version of the response tensor, by means of shell averaging over tensor trace: 
\begin{equation} 
\oint \HC_{ii}(\kappab,\tau) dS(\kappa) = 8\pi\kappa^2\Gav(\kappa,\tau).
\end{equation}

When measuring $\Gav(\kappa,\tau)$, a proper spatial and temporal discretization must be adopted. While in a DNS the discretizazion in $\kappa$ is easily derived from the Fourier representation of the velocity field (as for the energy spectrum $E(\kappa)$, see Ref. \cite{lamorgese-caughey-pope-2005}), the definition of the $\tau$-step is less obvious. Both the time resolution of the white-noise delta correlation, $\Delta \tau_w$, i.e. the time interval between successive updates of the random numbers, and the averaging time $T_{av}$, i.e. the time interval over which statistics are computed, must be chosen such that $\Gav$ is properly described and at the same time the computational requirements of the numerical simulation are kept reasonable. If $\tau_{min}$ indicates the smallest time scale at which proper convergence of the response is sought, $\Delta\tau_w$ must be chosen so that $\Delta \tau_w \leq \tau_{min}$. Assuming uniform sampling of the response in $N_c$ time instants separated by $\Delta \tau$, the time horizon available to represent the decay of the entire response must be greater than the whole response decay time $\tau_{max}$ at the lower wave number in the range of interest, i.e. $N_c \Delta \tau \ge \tau_{max}$. Proper convergence of the average response obviously requires $T_{av}/\tau_{max} \gg 1$. Indeed, while $\kappa_d$ controls $\Delta\tau$ resolution and then $\Delta t$, i.e. the time integration step, the largest inertial wave number dictates $N_c$ and $T_{av}$. 

Given such contrasting requirements, characterizing the function $\Gav(\kappa,\tau)$ in the whole universal range of scales via a sole measurement is possible but computationally demanding. Hence the entire function $\Gav(\kappa,\tau)$ can be measured through more than one uniform $\tau$-grid, so that the response is probed within several sub-ranges of scales, leveraging their reduced extent. $\Gav(\kappa,\tau)$ is then measured in a wide range of scales via a limited number of DNS runs, each of which requires roughly the same computational effort. These simulations are independent, and can be run simultaneously if the available computing power allows. However, for the results to obey the linearity costraint, the level of the introduced ``noise energy'', $\epsilon \Delta \tau_w$, must be kept constant across the different $\Delta \tau$ resolutions adopted at different scales. This means that a larger $\Delta \tau$ implies a reduced noise amplitude $\epsilon$ and a longer averaging time. This is partially compensated by the larger time step size allowed by the time resolution of the response at lower wave numbers.

%------------------------------------------------------------------------
\subsection{A test case: the purely viscous Stokes' response}
\label{sec:stokes}

The Stokes or viscous response represents the zero-order term in the expansion series of $\Gav$ as introduced in the context of renormalized perturbations, see Ref. \cite{leslie-1973,mccomb-1990}. The Stokes response, $\Gav^{(0)}$, can be easily derived from Eq. (\ref{eq:ist-resp}) after removal of the non-linear terms, thus providing the solution for pure viscous dynamics of 
the velocity field. Its analytical form reads:
\begin{equation}\label{eq:resp-stokes}
 \Gav^{(0)}(\kappa,\tau) = \exp(-\nu\kappa^2\tau).
\end{equation}

It is important to notice that the Stokes response has a deterministic nature, owing to the linearity of the Stokes operator: Kraichnan usually refers to it as ``statistically sharp''. The exact Stokes solution provides an useful tool for the validation of the full measurement procedure. To this purpose, the Stokes response can be retrieved from a DNS of the fully non-linear NSE through a numerical linearization. In this way the algorithm employed for the measurement in the turbulent case is exactly that previously described in \S\ref{sec:technique}, but a null initial condition is adopted, the energy-driving forcing of Eq. (\ref{eq:forcing}) is turned off, and only the white-noise perturbation is applied. If $\epsilon \ll 1$, no evolution toward turbulence dynamics is produced, and non-linear terms $\OC(\epsilon^2)$ can be neglected with respect to the linear ones $\mathcal{O}(\epsilon)$ defining the Stokes equation.

\begin{figure}
\centering
\includegraphics[width=0.45\textwidth]{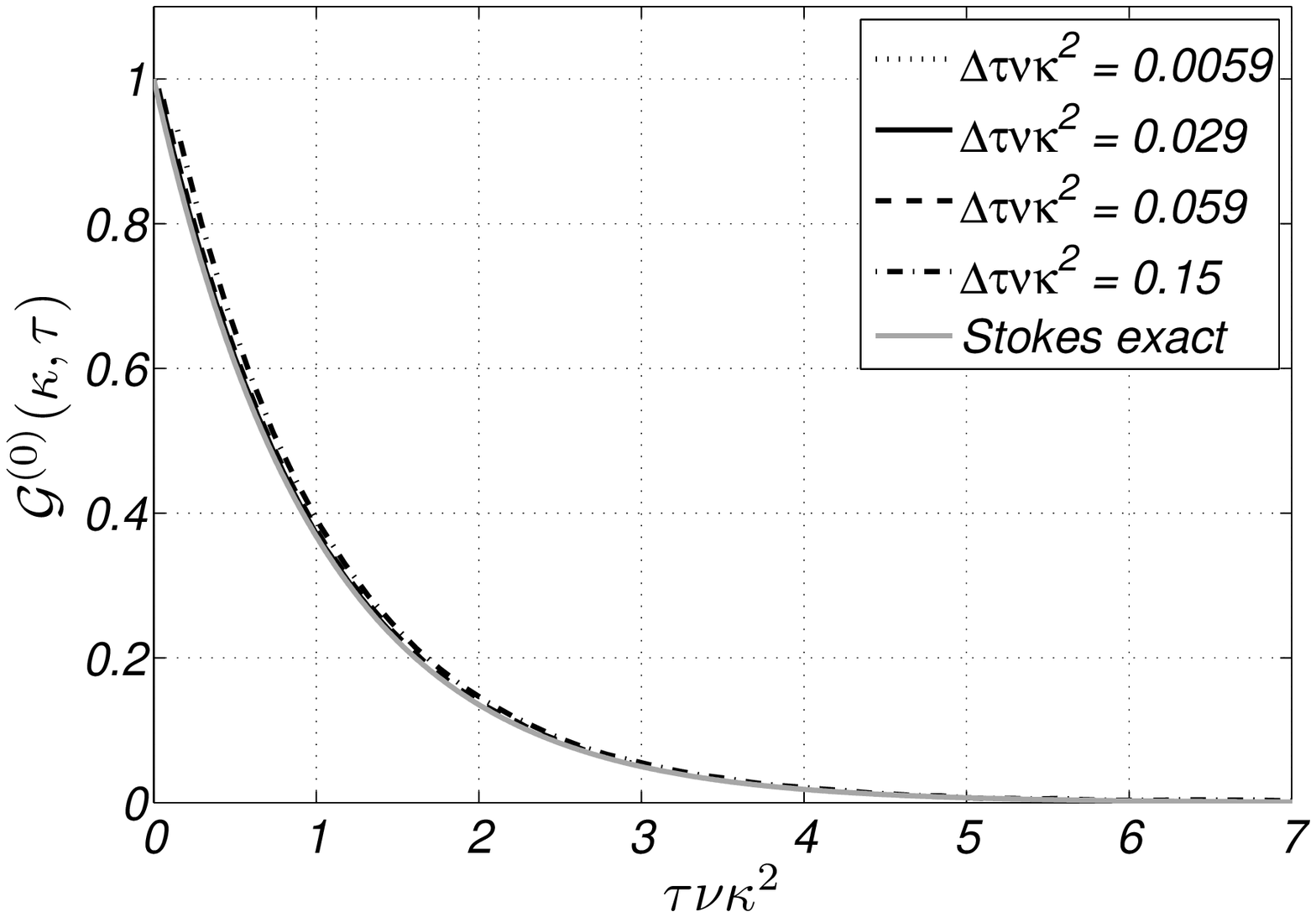}\\
\includegraphics[width=0.45\textwidth]{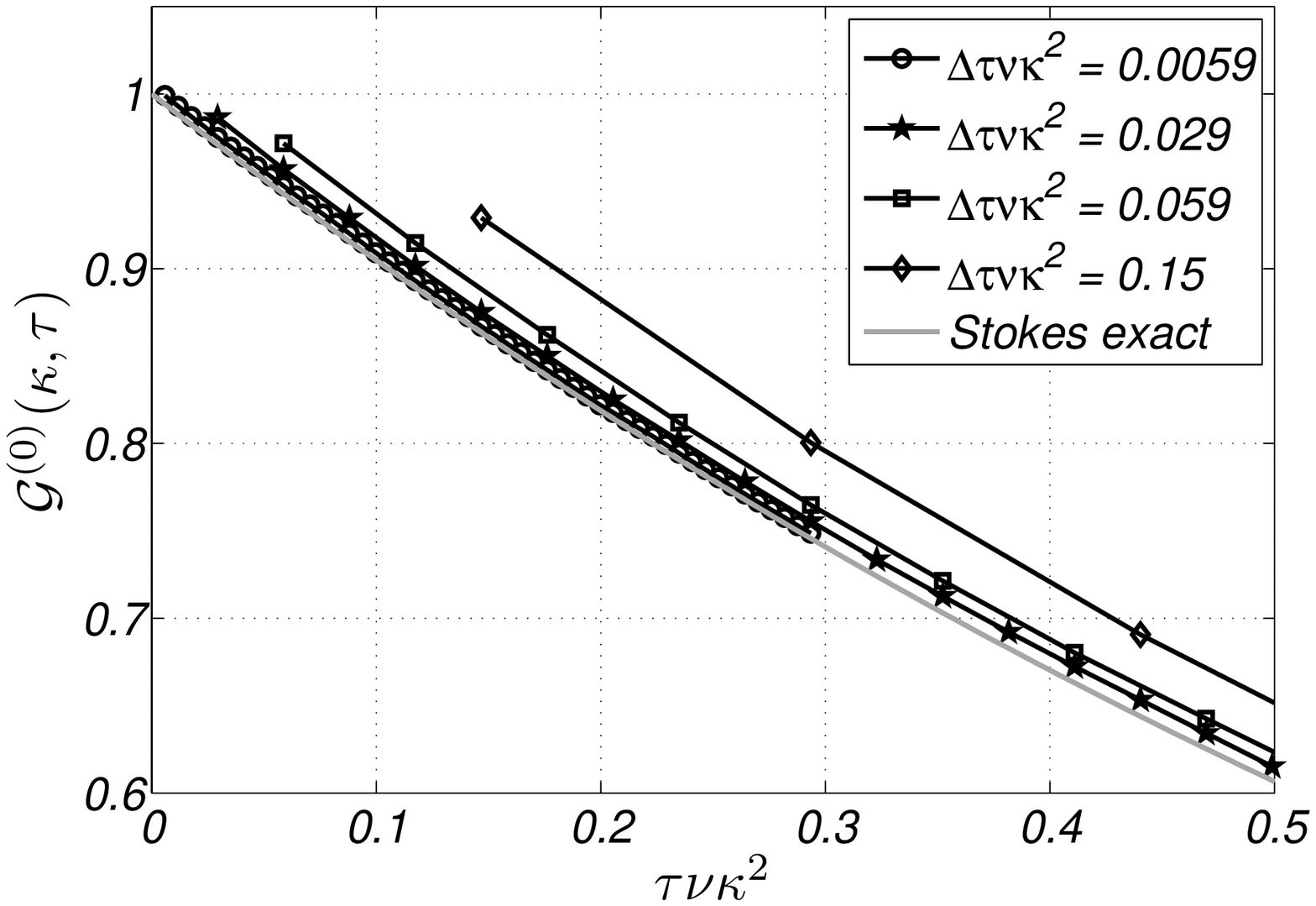}\\
\includegraphics[width=0.45\textwidth]{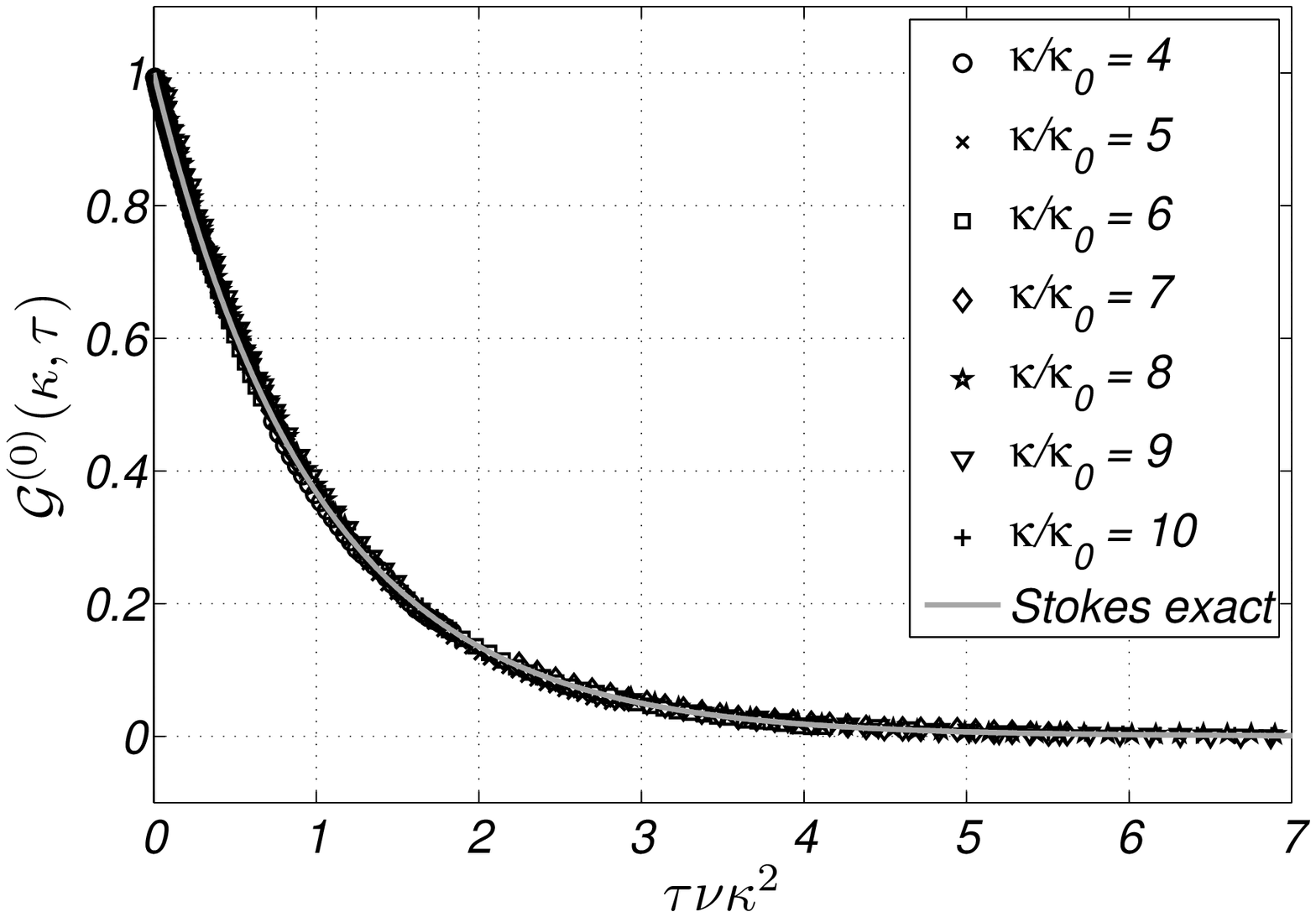}
\caption{Time decay of the measured Stokes response $\Gav^{(0)}$ at different $\Delta\tau$ with $\epsilon = 0.001$ and $T_{av}\nu\kappa^2 = 733.76$. Top: comparison between the measured and exact $\Gav^{(0)}$ at fixed $\kappa/\kappa_0 = 8$. Center: zoom of top plot for $\tau\nu\kappa^2 \ll 1$. Bottom: Stokes response at several wave number $\kappa/\kappa_0$ vs. non-dimensional time separation, emphasizing collapse with local viscous time scaling $\tau \nu \kappa^2$.}
\label{fig:stokes1D}
\end{figure}

The Stokes response has been measured in numerical experiments with a spatial resolution of $32^3$ modes (before dealiasing) and $N_c=50$. In these simulations $\Delta \tau = \Delta \tau_w$ is adopted, and several values of $\Delta \tau_w$ are used to investigate time resolution effects. In Fig. \ref{fig:stokes1D} (top and center) the time decay of the Stokes response at $\kappa / \kappa_0 = 8$ is plotted. The exact solution and the measured one agree very well (top figure). At small viscous time separations, $\tau\nu\kappa^2 < 1$, proper converge of the measured response towards the exact one is observed (center figure) to depend on the $\Delta \tau_w$ resolution. Convergence of the response for different noise amplitude $\epsilon$ and averaging time $T_{av}$ has been additionally verified (not shown). Lastly, the measured Stokes response plotted at different wave numbers (bottom figure) is observed to possess the expected collapse when local viscous time scaling is adopted for $\tau$. 

%%%%%%%%%%%%%%%%%
\section{The turbulent response}
\label{sec:results}

\begin{table}
\caption{Parameters for the DNS of HIT carried out in the present work.}
\label{tab:dns} 
\begin{ruledtabular}
\begin{tabular}{cccccccc}
  $N$ & $\kappa_{max}/\kappa_0$ & $\kappa_d/\kappa_0$ & $P$ & $\kappa_f/\kappa_0$ & 
  $Re=\left(\dsfrac{\kappa_d}{\kappa_f}\right)^{4/3}$ & $Re_\lambda$ &
  $u_0\left(\dsfrac{\kappa_f}{P}\right)^{1/3}$ \\
  128 &  42  & 28 & 1 & 3 & 20 & 55 & 1.7862\\
  192 &  63  & 42 & 1 & 3 & 34 & 77 & 1.8453\\
  256 &  84  & 56 & 1 & 3 & 49.5 & 94 & 1.8611\\
\end{tabular}
\end{ruledtabular}
\end{table}

\begin{table}
\caption{Discretization parameters for the DNS-based measurement of the response function.}
\label{tab:resp}
\begin{ruledtabular}
 \begin{tabular}{ccccccc}
  $N$ & $Re_\lambda$ & \textit{Run} & $N_c$ & $\Delta\tau u_0\kappa_d$ & $\epsilon$ & $T_{Av}u_0\kappa_d$ \\
\hline
 \multirow{2}*{128} & \multirow{2}*{55} &  1  &  150  & 0.05202 & 0.00093 & 1.3004e5 \\
                     &                  &  2  &  150  & 0.0322 & 0.0015 &  5643.7\\
\hline
  \multirow{2}*{192} & \multirow{2}*{77} &  1  &  150  & 0.0484 & 6.667e-4 & 9680 \\
                     &                   &  2  &  150  & 0.0322 & 0.001 & 5643.7\\
\hline
  \multirow{3}*{256} & \multirow{3}*{94} &  1  &  150  & 0.0614 & 4.3529e-4 & 9981.6\\
                     &                   &  2  &  150  & 0.0376 & 7.1154e-4 & 6580\\
                     &                   &  3  &  150  & 0.0267 & 0.001 & 2069.2\\  
\end{tabular}
\end{ruledtabular}
\end{table}

Several DNS have been run to measure the impulse response of homogeneous isotropic turbulence. They are grouped into simulations with $128^3$, $192^3$ and $256^3$ Fourier modes, before dealiasing. Table \ref{tab:dns} summarizes the discretization parameters of the simulations carried out without white-noise forcing, whereas Table \ref{tab:resp} lists all the simulations run to measure the response function, together with the values of the parameters used to discretize and measure $\Gav(\kappa,\tau)$. The attained values of $Re_\lambda$ are low or moderate, ranging from $Re_\lambda=55$ to $Re_\lambda=94$. Moreover, given our limited computational resources, the response is probed only at high wave-numbers. As explained in \S\ref{sec:technique}, long averaging times are in fact required for proper convergence of the mean reponse at low wavenumbers. However, it is important to emphasize here that the proposed measurement technique is capable of measuring the impulse response at any scale, provided that adequate computational resources are available.

In \S\ref{sec:lin} the linear behavior and the convergence of time averages are demonstrated, whereas in \S\ref{sec:scaling} the behavior of the response and its scaling within the dissipative universal subrange are investigated. The correlation function is also computed during the stationary reference HIT simulations, so that a comparison with the impulse response function in terms of the classical FDR will be given at least in the range of scales here considered.

%-----------------------------------------------
\subsection{Linearity and time average}
\label{sec:lin}

\begin{figure}
\centering
\includegraphics[width=0.45\textwidth]{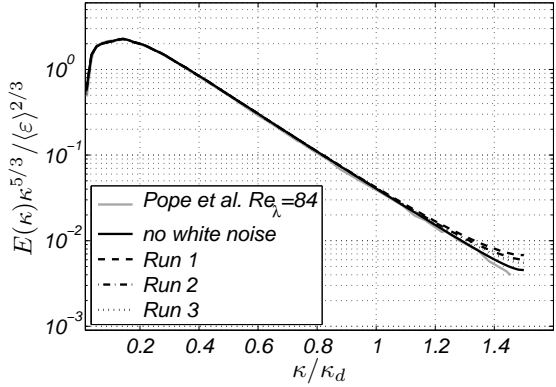}
\caption{Comparison between compensated energy spectra from HIT and from white-noise-forced HIT employed for measuring $\Gav(\kappa,\tau)$, $Re_\lambda=94$ ($N=256$) (cfr. Table \ref{tab:resp}). Note the absence of white-noise effects for $\kappa / \kappa_d < 1$.}
\label{fig:noisy-spectra}
\end{figure}

A key issue when measuring the response function $\Gav(\kappa,\tau)$ is the proper choice of amplitude $\epsilon$ for the white-noise forcing. Indeed the true turbulent impulse response reduces to its linear counterpart $\Gav$ only for vanishing ``noise energy'', i.e. when $\epsilon \Delta \tau_w \rightarrow 0$. In a finite setting, a reasonable preliminary requirement is that the white-noise forcing does not affect turbulence statistics appreciably. Then, suitable convergence of the measured response to $\Gav$ is observed when the function $\Gav / \epsilon$ becomes indipendent on $\epsilon$. Indeed, as discussed in \S\ref{sec:technique}, given a time resolution $\Delta \tau_w$, $\epsilon$ represents the linearity control parameter. Since the white-noise forcing is spatially distributed over all the scales, the linearity threshold is fixed by pertubations effects on smallest scale dynamics, i.e. the viscous scales: when looking for the response at lower wave numbers with larger $\Delta \tau = \Delta \tau_w$, $\epsilon$ must be reduced to preserve the linear response at higher wave numbers. 
 
Fig. \ref{fig:noisy-spectra} provides a comparison between the energy spectrum $E(\kappa)$ computed in standard HIT DNS and those from simulations forced with white noise at $Re_\lambda=94$. Analogous results (not shown here) holds for the other spatial resolutions listed in Table \ref{tab:resp}. The value of $\epsilon$, that should be maximized in order to increase S/N and hence to reduce the required averaging time, is chosen so that marginal effects on the spectrum are confined within the numerical wavenumbers larger than the Kolmogorov scale, $\kappa > \kappa_d$. Moreover, table \ref{tab:stat_dev} quantifies the little variations in statistics like $\av{\varepsilon}$ and $\av{k}$ due to white-noise forcing: $\Delta \av{\varepsilon}$ and $\Delta \av{k}$ are less then $0.4\%$ and $1.9\%$ respectively. $\Delta \av{\varepsilon}$, that is computed with respect to the exact asymptotic value $P$, is of the same order of the variations of $\av{\varepsilon}$ observed in different runs of standard HIT DNS. $\Delta \av{k}$, that is computed against the value of $\av{k}$ obtained from standard HIT DNS (Run 0), seems to be relatively larger. However, it should be recalled how the accurate convergence of this statistic, that belongs to large scales, requires averaging over many turnover times. Indeed, the observed $\Delta \av{k}$ are of the same order of $\av{k}$ fluctuations in standard HIT DNS, under the feedback action of the energy-driving forcing scheme with an averaging time of $T_{av}(P\kappa_f^2)^{1/3} \approx 250$, to which standard HIT averaged values are referred.

\begin{table}
\caption{Effect of the white-noise forcing on $\av{\varepsilon}$ and $\av{k}$, for various spatial resolutions. The reference simulations without white noise are indicated as ``Run 0''.}
\label{tab:stat_dev}
\begin{ruledtabular}
\begin{tabular}{ccccccc}
  $N$ & $Re_\lambda$ & \textit{Run}  & $\av{\varepsilon}/P$ & $\av{k}/(P/\kappa_f)^{2/3}$ & $\Delta\av{\varepsilon}\%$ 
  & $\Delta\av{k}\%$\\
 \multirow{3}*{128} & \multirow{3}*{55} &  0  & 0.999595 & 4.7856 & 0.0405 & - \\
                     &                  &  1  & 1.00057  & 4.7904 & 0.057 &  0.1\\
                     &                  &  2  & 1.00056  & 4.7374 & 0.056 &  1.007\\
  \multirow{3}*{192} & \multirow{3}*{77} &  0  & 0.999788 & 5.0538 & 0.0212 & - \\
                     &                   &  1  & 1.00177  & 4.9612 & 0.177 & 1.83 \\
                     &                   &  2  & 1.0025   & 5.0063 & 0.25 & 0.94 \\
  \multirow{4}*{256} & \multirow{4}*{94} &  0  &  1.00155  & 5.1958 & 0.155 &  -    \\
                     &                  &  1   &  1.00268  & 5.1235 & 0.113 & 1.392 \\
                     &                  &  2   &  1.0007   & 5.1082 & 0.085 & 1.686 \\
                     &                  &  3   &  1.00354  & 5.2561 & 0.354 & 1.161 \\  
\end{tabular}
\end{ruledtabular}
\end{table}

\begin{figure}
\centering
\includegraphics[width=0.45\textwidth]{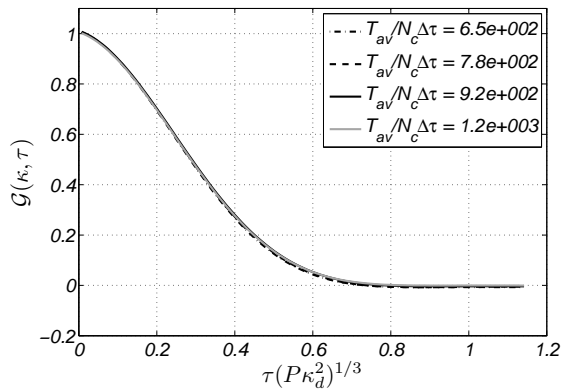}
\caption{Convergence of the measured $\Gav(\kappa,\tau)$ with respect to the averaging time $T_{av}$.  Results for Run 2 at $Re_\lambda=94$ (cfr. Table \ref{tab:resp}) are shown at rapresentative wavenumbers $\kappa/\kappa_d=0.75$.}
\label{fig:resp_Tav_conv}
\end{figure}

The convergence of the measured $\Gav(\kappa,\tau)$ with respect to the averaging time $T_{av}$ is also verified. In Fig. \ref{fig:resp_Tav_conv} responses measured with increasingly larger values of $T_{av}$ are shown for Run 2 at $Re_\lambda=94$. Adequate convergence is obtained at representative wavenumbers $\kappa/\kappa_d=0.75$ when $T_{av}/(N_c\Delta\tau) > 920$. At larger averaging times, the response curves become indistinguishable. A similar behavior has been verified for the other numerical experiments reported in Table \ref{tab:resp}. 

%-----------------------------------------------
\subsection{The response function and its scaling in the viscous universal subrange}
\label{sec:scaling}

\begin{figure}
\centering
\includegraphics[width=0.45\textwidth]{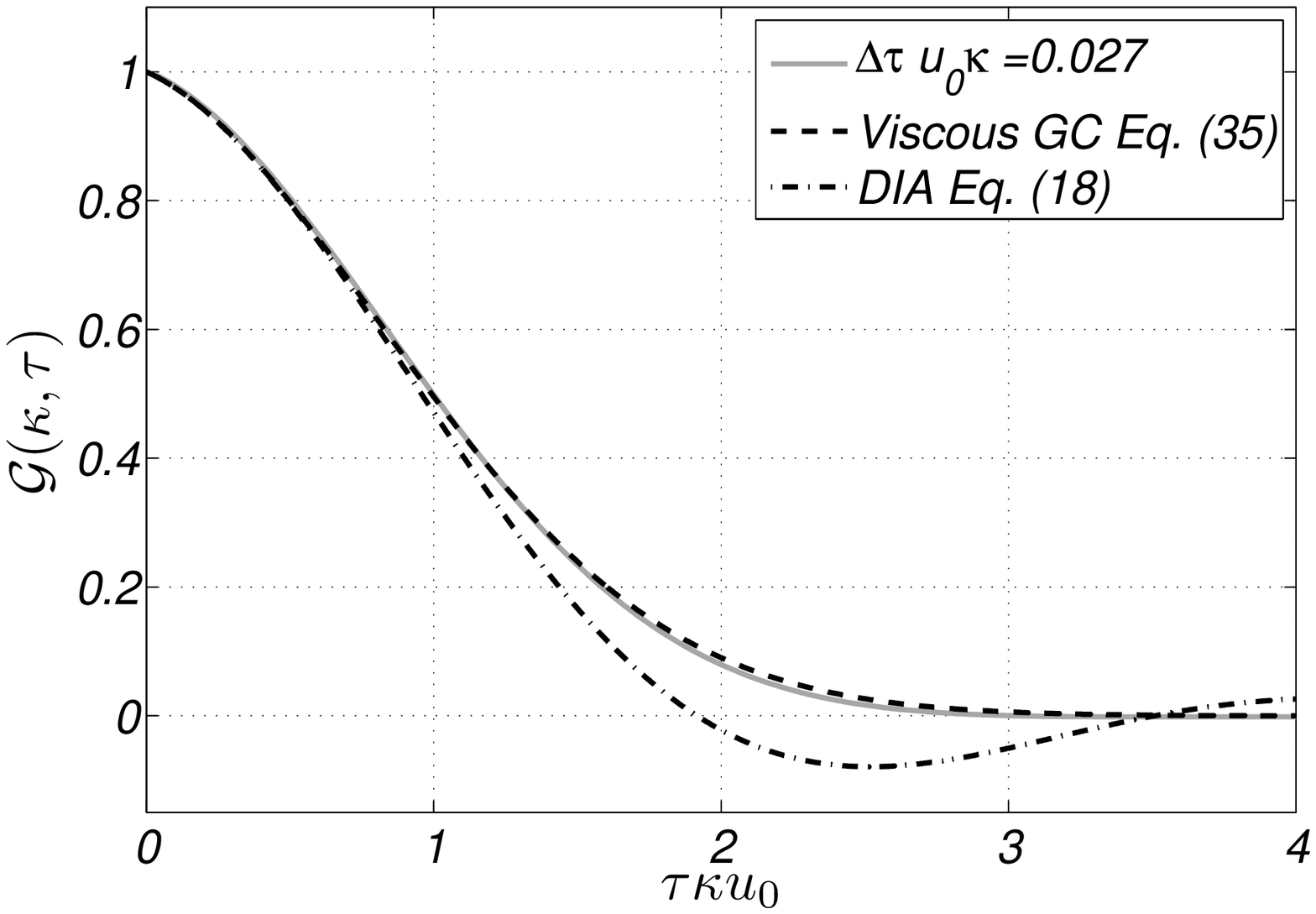}\\
\includegraphics[width=0.45\textwidth]{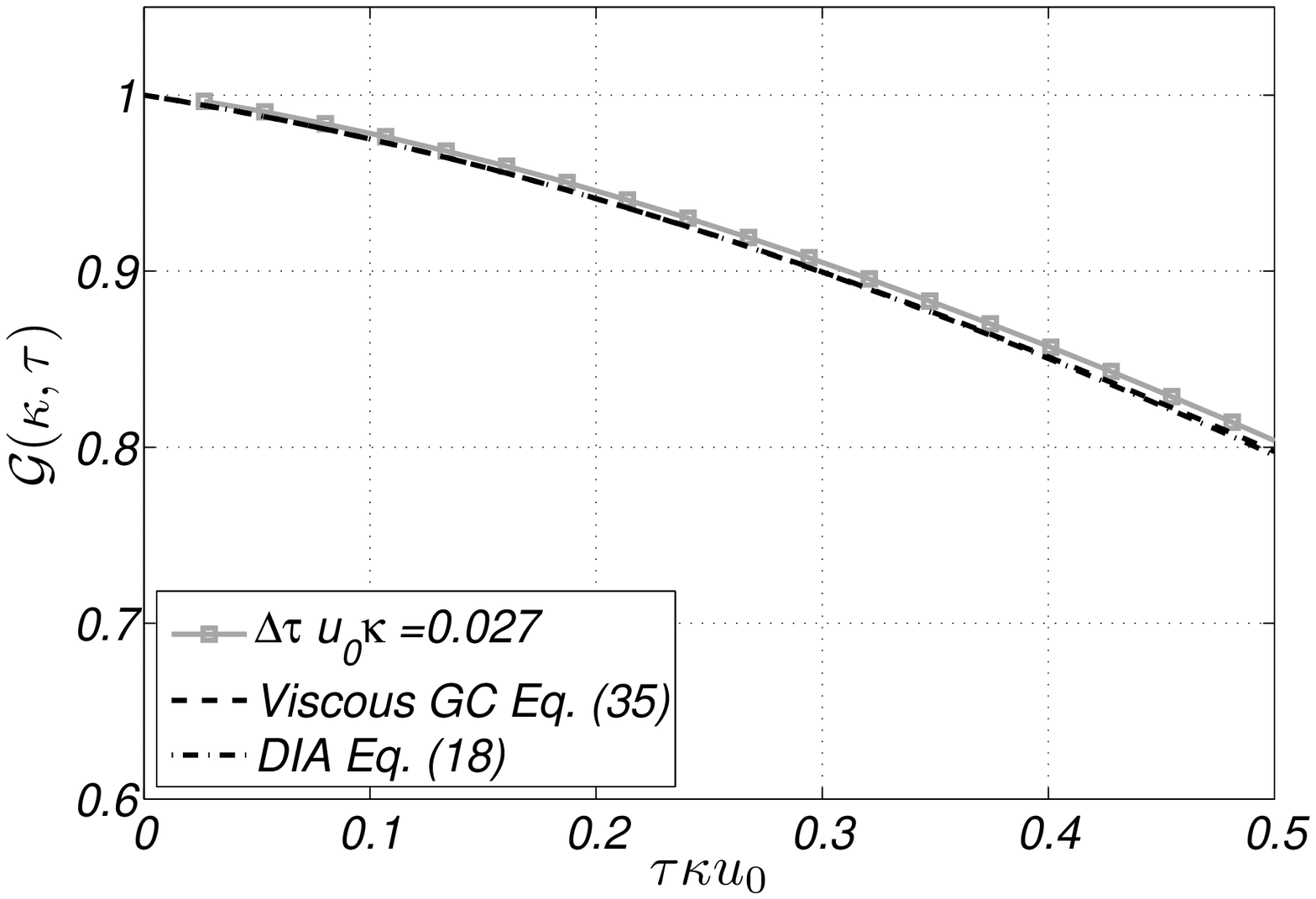}\\
\includegraphics[width=0.45\textwidth]{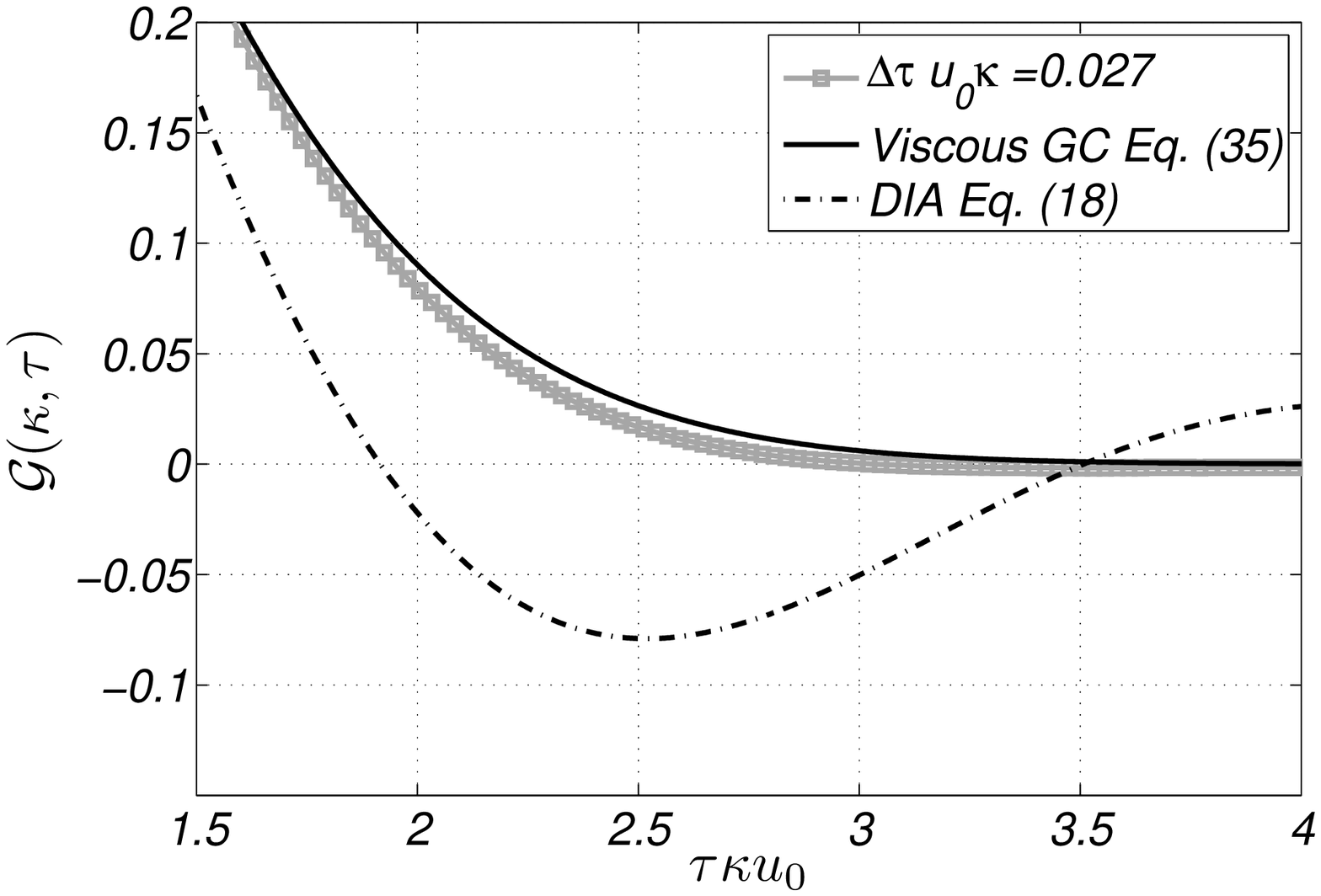}
\caption{Measured response function from Run 3 of Table \ref{tab:resp} at $Re_\lambda=94$. Top: time decay of the response function at the Kolmogorov scale ($\kappa / \kappa_d=1$), compared with the DIA solution Eq. (\ref{eq:resp-dia}) and the viscous Gaussian-convective solution Eq. (\ref{eq:resp-vgc}). Center: zoom of the top plot for $\tau \kappa u_0 \ll 1$. Bottom: zoom of the top plot at large $\tau \kappa u_0$.}
\label{fig:NS_resp_1D_keta}
\end{figure}

The response function measured via the procedure illustrated above is first compared with its available analytical approximations, as given in the original DIA theory, see Ref. \cite{kraichnan-1959}:
\begin{equation}
\label{eq:resp-dia}
 \Gav(\kappa,\tau) = \exp(-\nu\kappa^2\tau)\dsfrac{J_1(2u_0\kappa\tau)}{u_0\kappa\tau},
\end{equation}
and in the analysis of random convection effects \cite{kraichnan-1964-b,mccomb-1990} from which the viscous Gaussian-convective response $\Gav_{GC}(\kappa,\tau)$ can be introduced: 
\begin{equation}\label{eq:resp-vgc}
 \Gav_{GC}(\kappa,\tau) = \exp(-\nu\kappa^2\tau-\dsfrac{1}{2} u_0^2\kappa^2 \tau^2), \quad \mbox{ with } \tau > 0.
\end{equation}
In both the previous equations $u_0$ represents the r.m.s. value of turbulent fluctuations and it can be easily recognized the Stokes term, Eq. (\ref{eq:resp-stokes}),  which reflects the viscous response of the corresponding linear operator. It is important to recall that while the non-viscous term of Eq. (\ref{eq:resp-dia}) is derived as an approximated solution to the DIA equations, the corrisponding one of Eq. (\ref{eq:resp-vgc}) empirically follows from the analogy with the solution of the idealized problem of \textit{pure random convection} introduced by Kraichnan in Ref. \cite{kraichnan-1964-b} with the Random Galilean Invariance (RGI) postulate to explain the failure of DIA in yielding a Kolmogorov inertial-range scaling. Refs.  \cite{mccomb-shanmugasundaram-hutchinson-1989,mccomb-2005} give a more recent investigation on the role of random convection effects and RGI in renormalized perturbation expansions of the NSE.

A comparative view of these three response functions at $\kappa/\kappa_d=1$ is provided in Fig. \ref{fig:NS_resp_1D_keta} for $Re_\lambda=94$, Run 3. At time separations smaller than the local energy time scale, i.e. for $\tau u_0 \kappa < 1$, the true measured response is in good agreement with the DIA response function and the viscous Gaussian-convective solution. The latter result does not come as a surprise: even though the turbulent field is definitely non-Gaussian, at times smaller than the characteristic correlation time the Gaussian approximation still applies, see \cite{leslie-1973}. The unexpected result, however, is that the Gaussian convective solution still approximates very well the measured response at larger times, whereas the DIA solution clearly deviates from it. 

\begin{figure}
\centering
\includegraphics[width=0.45\textwidth]{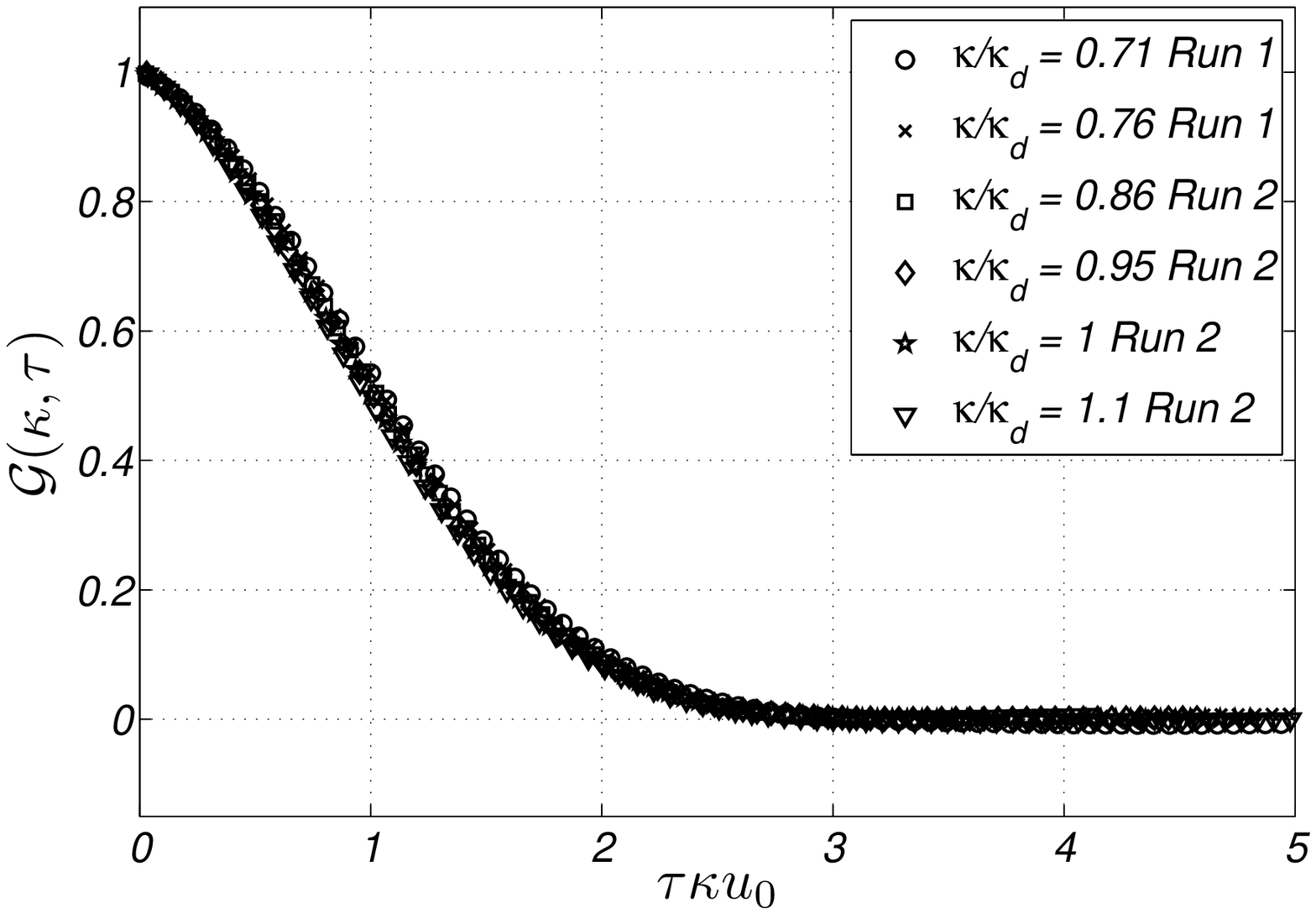}\\
\includegraphics[width=0.45\textwidth]{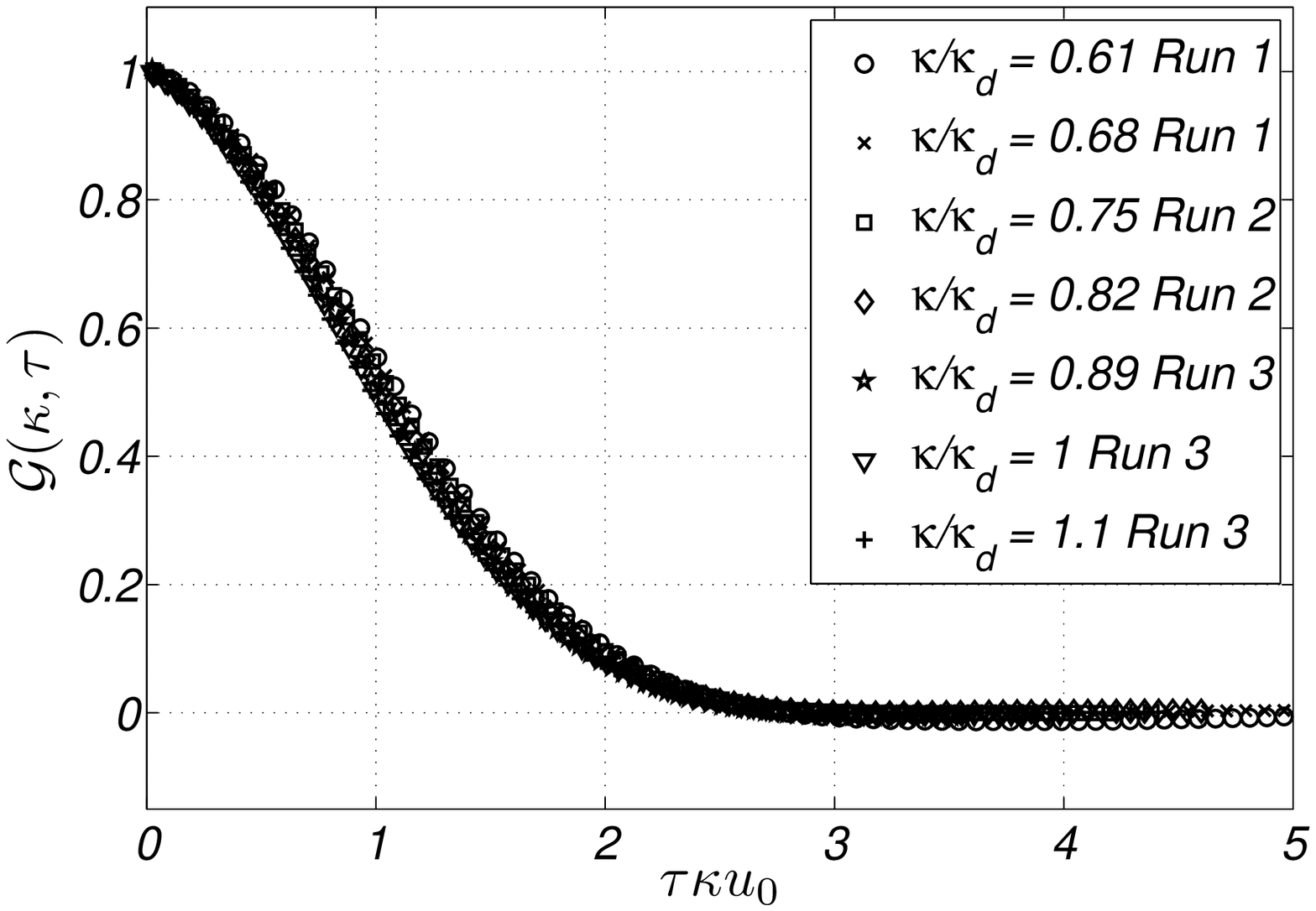}
\caption{Measured response functions plotted with convective scaling at different values of $Re_\lambda$. The responses are plotted versus non-dimensional time separation $\tau u_0 \kappa$, at several wavenumbers within the universal dissipative subrange (cfr. Table \ref{tab:resp}). Top: $Re_\lambda=77$. Bottom: $Re_\lambda=94$. For better clarity response functions are plotted using one of every two of the $N_c$ values. Convective scaling produces a good collapse of the curves.}
\label{fig:NS_resp_1D_conv_scaled}
\end{figure}

This evidence provides further motivation for investigating the convective response scaling in the viscous universal subrange. Response functions rescaled accordingly are plotted in Fig. \ref{fig:NS_resp_1D_conv_scaled}. For the entire range of values of $Re_\lambda$ considered in the present work, the convective scaling of the response function is clearly assessed.

\begin{figure}
\centering
\includegraphics[width=0.45\textwidth]{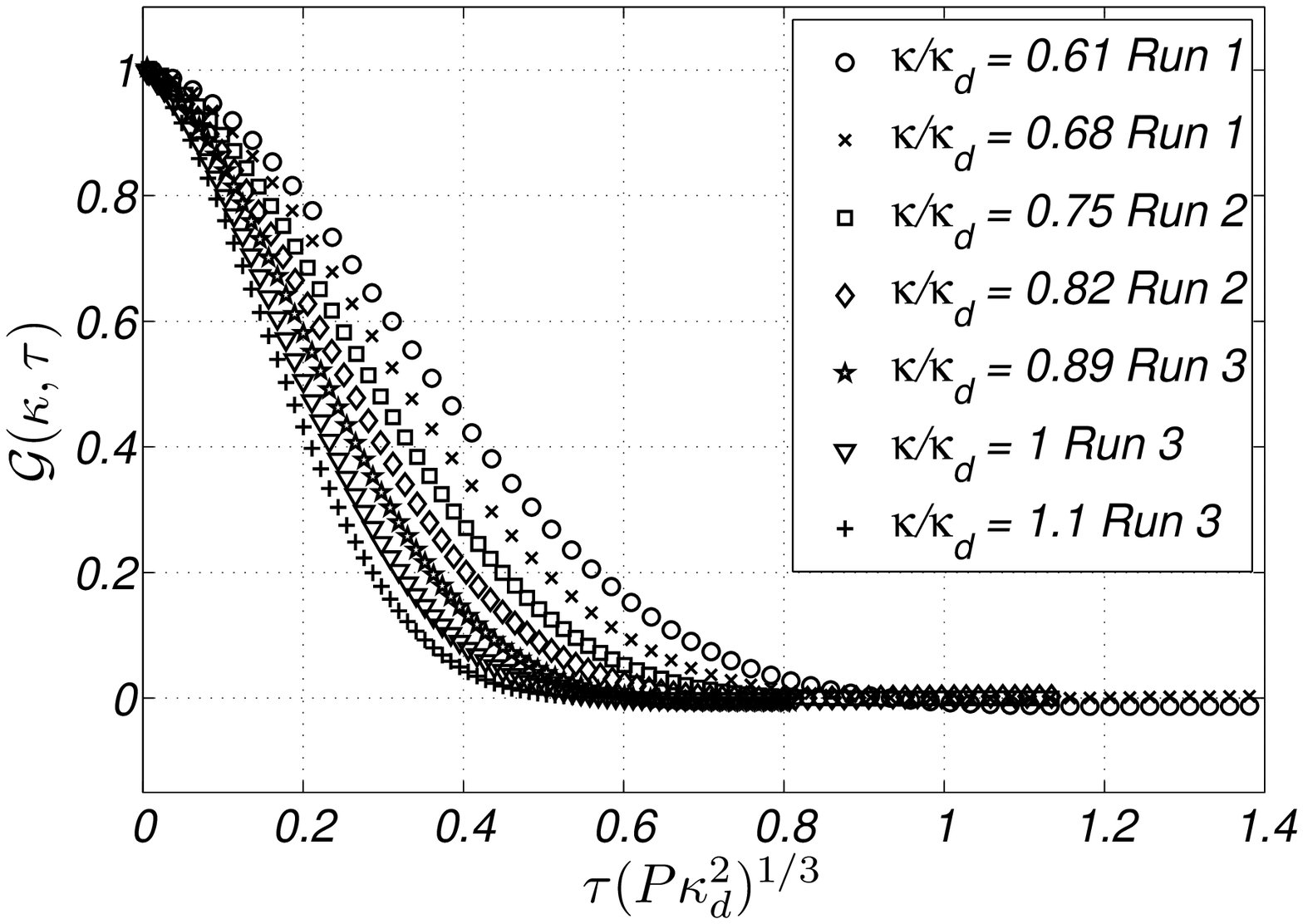}
\caption{Measured response functions with Kolmogorov scaling at $Re_\lambda=94$. The responses are plotted versus the non-dimensional time separation $\tau (P\kappa_d^2)^{1/3}$, at several wavenumbers within the universal dissipative subrange (cfr. Table \ref{tab:resp}). For better clarity response functions are plotted using one of every two of the $N_c$ values. Kolmogorov scaling is not successful in producing a collapse of the curves.}
\label{fig:NS_resp_1D_K41_scaled}
\end{figure}

To further support this statement, Fig. \ref{fig:NS_resp_1D_K41_scaled} shows the response functions tentatively plotted with Kolmogorov viscous scaling: it is evident that such scaling does not produce as good a collapse of the different curves when compared to the convective scaling employed in Fig. \ref{fig:NS_resp_1D_conv_scaled}.

%----------------------------------------------------------------------
\subsection{The correlation function and the FDR}

\begin{figure}
\centering
\includegraphics[width=0.45\textwidth]{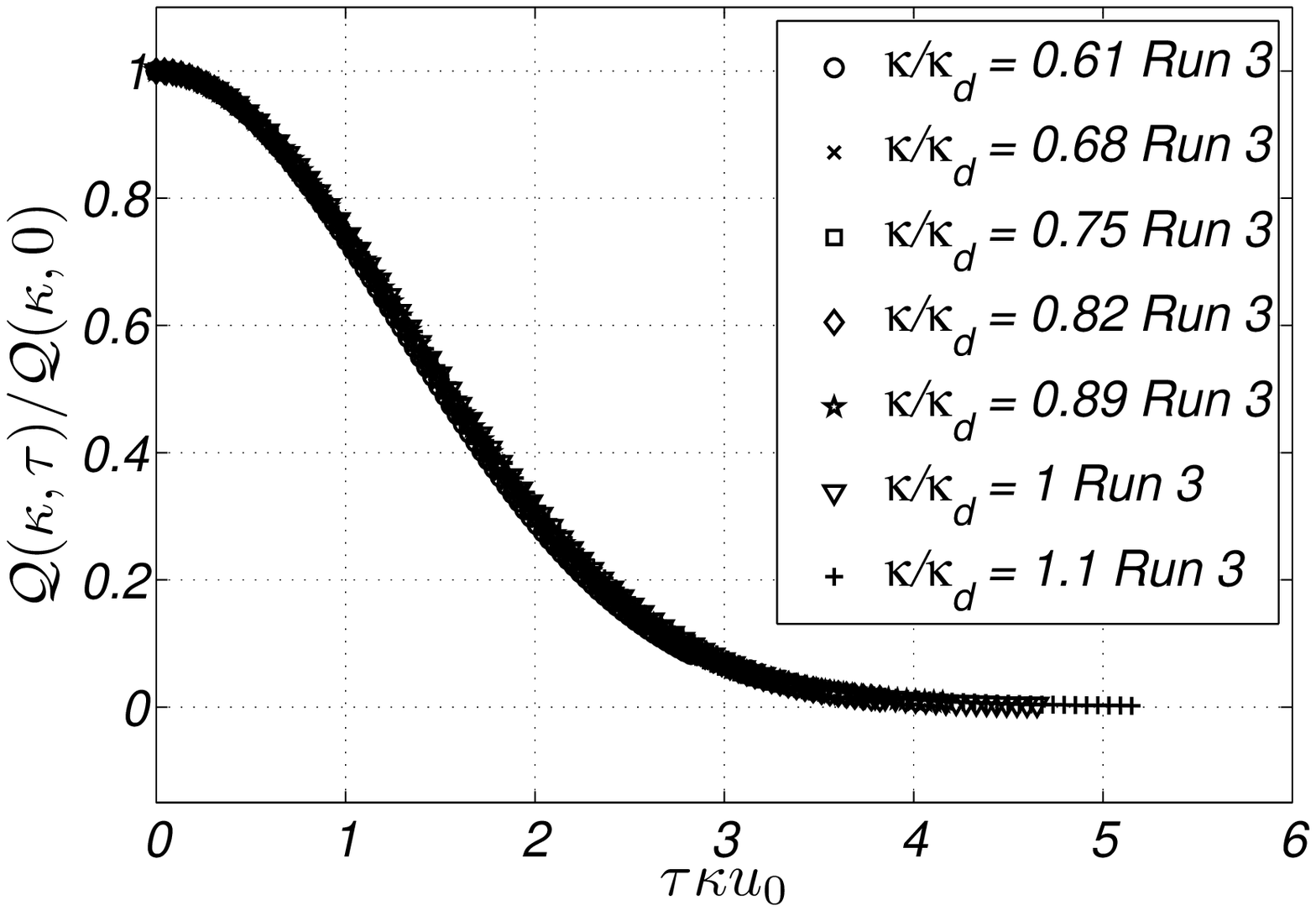}
\caption{Measured correlation function plotted with convective scaling. The normalized correlation function is plotted versus non-dimensional time separation $\tau u_0 \kappa$, for several wavenumbers in the universal dissipative subrange at $Re_\lambda=94$. Convective scaling produces a good collapse of the curves.}
\label{fig:NS_Q1D_conv_scaled}
\end{figure}

The mean correlation tensor is here introduced directly in its spectral form:
\begin{equation}
 \Qav_{ij}(\kappab,t,t') = \av{u_i(\kappab,t)u_j(-\kappab,t')},
\end{equation}
which reduces to the scalar function $\Qav(\kappa,\tau)$ in the homogeneous, isotropic stationary case:
\begin{equation}
 \Qav_{ij}(\kappab,t,t') = P_{ij}(\kappab)\Qav(\kappa,t-t').
\end{equation} 

\begin{figure}
\centering
\includegraphics[width=0.45\textwidth]{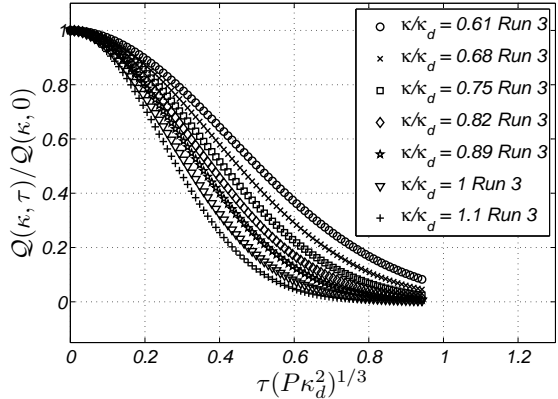}
\caption{Measured correlation function plotted with viscous scaling. The normalized correlation function is plotted versus non-dimensional time separation $\tau (P\kappa_d^2)^{1/3}$, for several wavenumbers in the universal dissipative subrange at $Re_\lambda=94$. Viscous scaling does not produce a collapse of the curves.}
\label{fig:NS_Q1D_K41_scaled}
\end{figure}

The correlation function $\Qav(\kappa,\tau)$ has been computed thanks to the DNS simulations carried out without white-noise forcing. At the various values of $Re_\lambda$ considered, the correspondent smallest $\Delta \tau$ time resolution employed for measuring the response function has been used. Discretization details can be found in Table \ref{tab:resp}. The number of time separations at which the correlations are stored is $N_c=200$ for the two cases respectively at $Re_\lambda=55$ and $Re_\lambda=77$, while $N_c=175$ has been employed for the case at $Re_\lambda=94$ due to memory limitations. Proper convergence of the results with respect to $T_{av}$ has been verified, according to what has been done for the response function itself. Similarly to what has been observed for $\Gav(\kappa,\tau)$, the scaling of the normalized correlation function, $\Qav(\kappa,\tau)/\Qav(\kappa,0)$, is captured by the local convective time scale $(\kappa u_0)^{-1}$ in the universal viscous subrange investigated here. This is illustrated for $Re_\lambda=94$ in Fig. \ref{fig:NS_Q1D_conv_scaled}, whereas the inadequacy of Kolmogorov viscous scaling is shown in Fig. \ref{fig:NS_Q1D_K41_scaled}. The same behavior can be also observed to hold for the correlations obtained at $Re_\lambda=55$ and $Re_\lambda=77$ (not shown here). 

The response and correlation functions, as measured from our DNS experiments, can be compared
through the well known FDR:
\begin{equation}
 \Qav(\kappa,\tau) = \Gav(\kappa,\tau)\Qav(\kappa,0).
 \label{eq:scalfdr}
\end{equation}

This relation has been originally derived in the context of Hamiltonian dynamical systems at equilibrium for which a canonical 
distribution holds. Only in the last decades, the applicability of the FDR to the wider class of non-linear chaotic dynamical systems 
has been addressed on a theoretical basis \cite{falcioni-isola-vulpiani-1990, falcioni-isola-vulpiani-1991, marconi-etal-2008}. 
For this class of systems (to which fluid turbulence belongs) a generalized FDR is demonstrated to hold, provided the system is 
{\em dynamically mixing}: only when a Gaussian distribution holds for the invariant probability distribution, the generalized FDR 
reduces to the classical form of Eq. (\ref{eq:scalfdr}). Obviously Eq. (\ref{eq:scalfdr}) cannot be exact for fully developed fluid 
turbulence, for which both experimental and numerical investigations have shown marked departures from 
Gaussianity, with long tails in the PDF and intermittent behavior.
However on an intuitive ground, one would expect a proportionality between the response and the correlation 
function to hold, at least in terms of characteristic time scales, respectively indicated by $\tau_{\Gav}(\kappa)$ and 
$\tau_{\Qav}(\kappa)$ \footnote{$\tau_{\Gav}(\kappa)$ and $\tau_{\Qav}(\kappa)$ are to be intended respectively as the separation 
time $\tau$ by which $\Gav(\kappa,\tau)$ and $\Qav(\kappa,\tau)$ reduce to the same percentage of their initial value.}. FDR has been then successfully applied in the context of climate study on sensitivity analysis with respect to external 
perturbations and parameters \cite{bell-1980, lacorata-vulpiani-2007}, as well as in viscosity renormalization 
\cite{carnevale-frederiksen-1983}. Nevertheless, in Refs. \cite{lacorata-vulpiani-2007, marconi-etal-2008} it is noted how in many such 
attempts the Gaussian form of the FDR has been often acritically invoked, with little awareness about its inherent limitations.
Moreover, in the field of spectral closures, different opinions exist on the possibility to recover the classical FDR in the Eulerian rather than in the Lagrangian framework. In Ref. \cite{kaneda-1981} the Gaussian form of the FDR is exactly recovered within the Lagragian renormalized approximation of turbulence, where the Lagrangian response function is introduced. In the Eulerian frame the proper use of the FDR has been recently addressed by Kiyani and McComb \cite{kiyani-mccomb-2004}. In their paper they show that FDR as stated in Eq. (\ref{eq:scalfdr}) is exact up to second order in renormalized perturbation expansions of NSE, hence it can be properly used in related closure formulations \cite{mccomb-kiyani-2005}. However, Kraichnan suggested \cite{kraichnan-2000} that even a valid Gaussian FDR would not immediately be a step forward in the closure problem. In Kraichnan's view, the strong departure from equipartition in the inertial range is not followed by a corresponding strong violation of the Gaussian FDR in the Eulerian frame. This is because large-scale random convection dominates the decay of both the response and the correlation functions, with corresponding time scales for mode $\kappa$ ruled by the local characteristic convective time $(\kappa u_0)^{-1}$. Kraichanan's analysis thus implies that the local dynamics cannot be captured by the elementary FDR, and the expected deviations can be found only by looking to a generalized FDR, that involves a Lagrangian form of the statistics.

\begin{figure}
\centering
\includegraphics[width=0.45\textwidth]{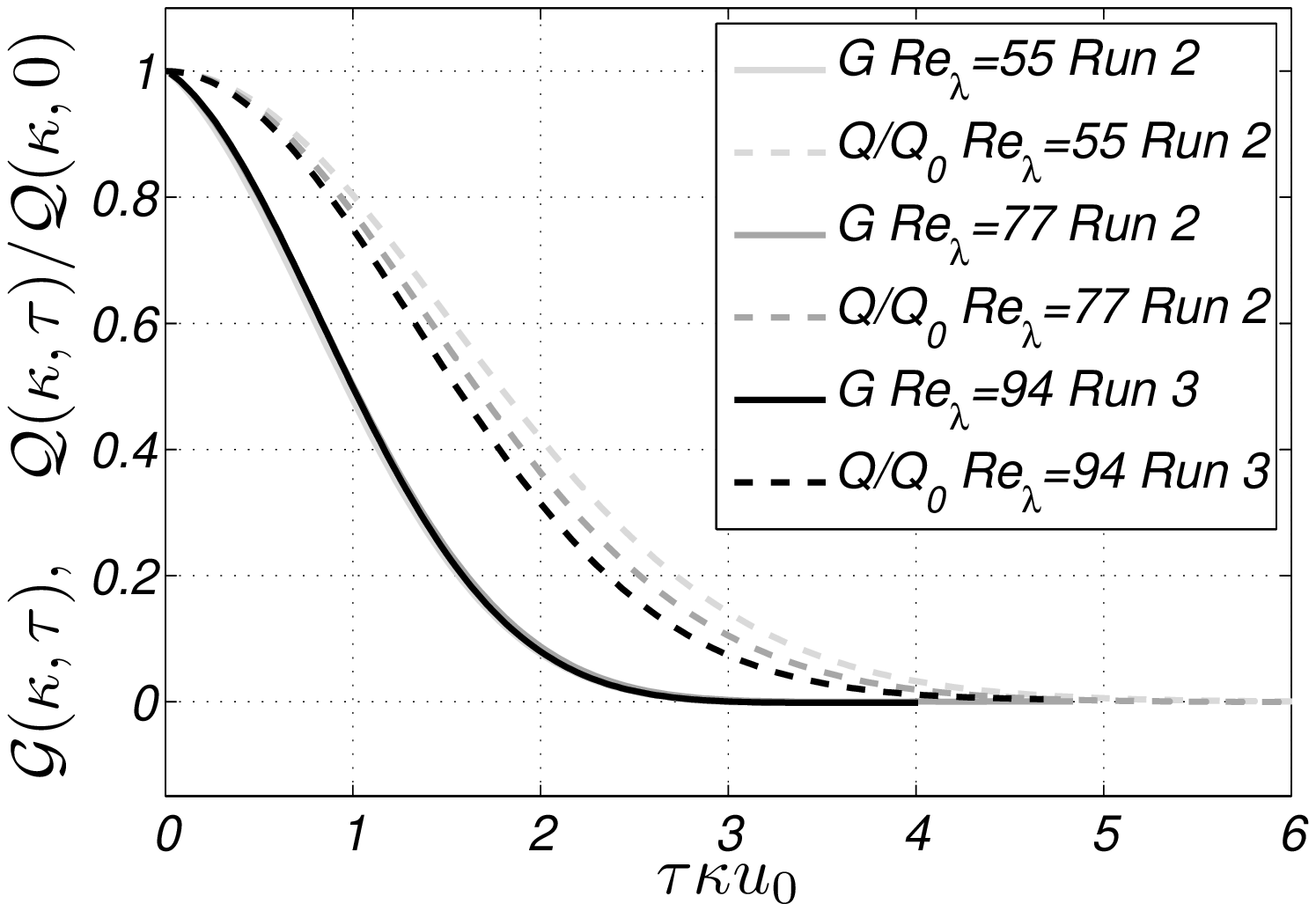}\\
\includegraphics[width=0.45\textwidth]{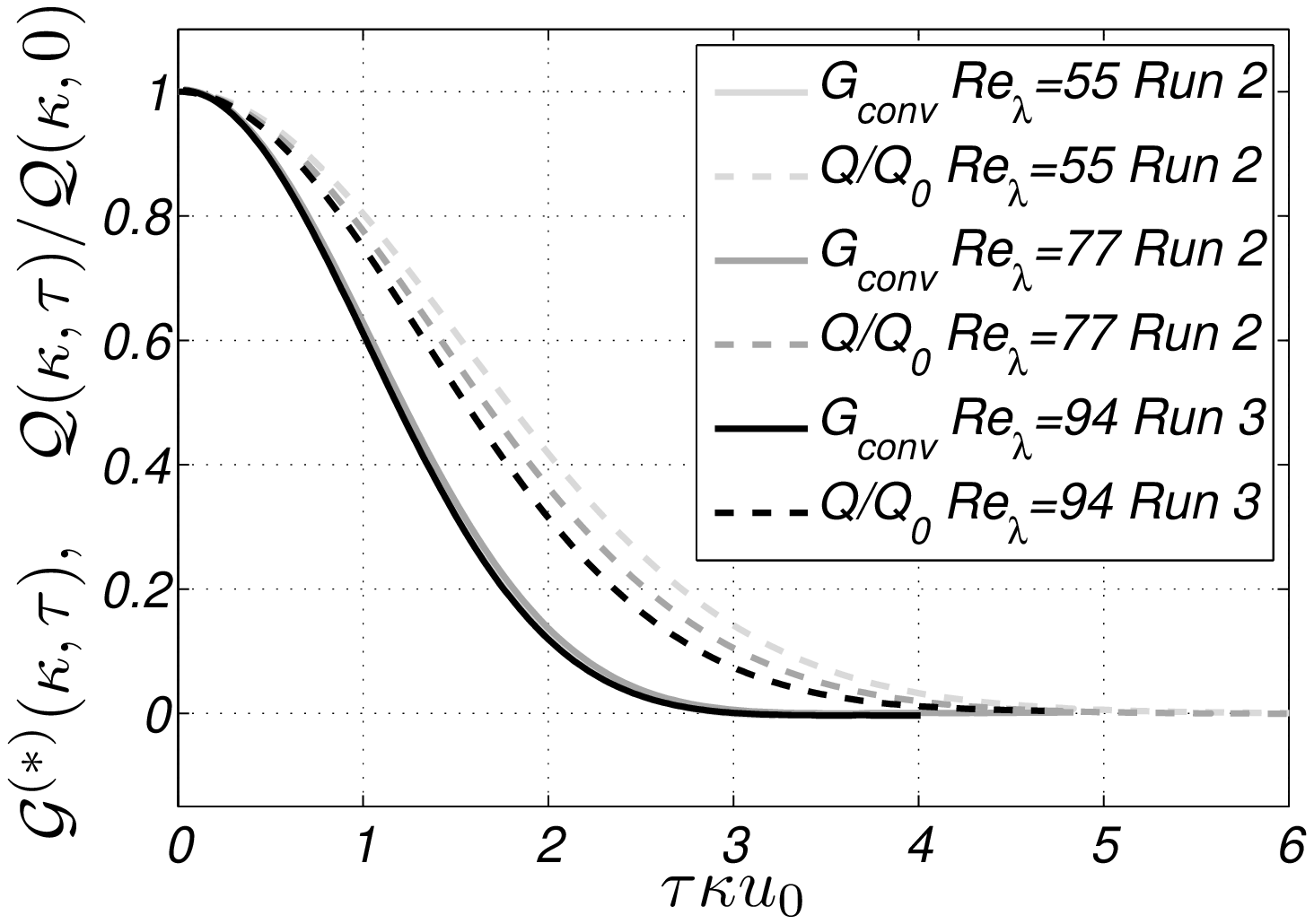}\\
\includegraphics[width=0.45\textwidth]{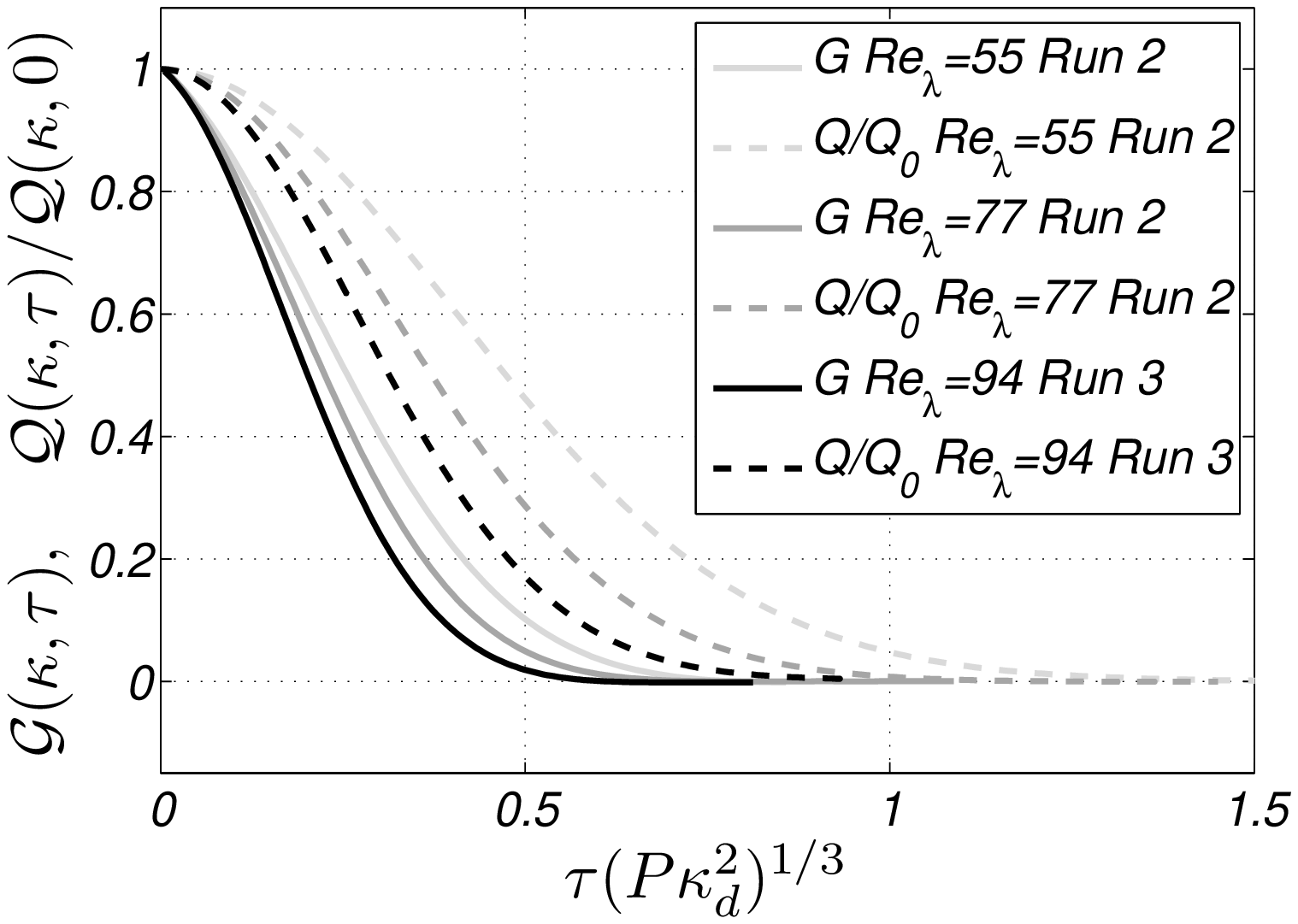}
\caption{Measured response functions $\Gav(\kappa,\tau)$ (continuous lines) and measured normalized correlation functions $\Qav(\kappa,\tau)/\Qav(\kappa,0)$ (dashed lines) at the Kolmogorov scale $\kappa=\kappa_d$ for several values of $Re_\lambda$. Top: convective scaling. Center: convective scaling, but using only turbulent-diffusive part of the response is used. Bottom: viscous Kolmogorov scaling.}
\label{fig:NS_FDR_kd}
\end{figure}

These considerations motivate investigating the approximation introduced by the Gaussian FDR within the Eulerian frame: a preliminary assessment is given Fig. \ref{fig:NS_FDR_kd} where $\Gav(\kappa,\tau)$ and $\Qav(\kappa,\tau)/\Qav(\kappa,0)$ are plotted together 
for $\kappa$ fixed at the Kolmogorov scale, i.e. for $\kappa / \kappa_d=1$.
As expected from theoretical arguments, see for example Ref. \cite{leslie-1973}, a longer decorrelation time is observed for 
$\Qav(\kappa,\tau)/\Qav(\kappa,0)$ when compared to $\Gav(\kappa,\tau)$. However the time scales $\tau_\Gav(\kappa)$ and 
$\tau_\Qav(\kappa)$ turn out to be of the same order. At fixed $Re_\lambda$, the plots in convective units of 
Fig. \ref{fig:NS_FDR_kd} (top and center) suggest that the response and the correlation functions are strictly related within the 
whole dissipative subrange of scales, owing to their inherent energy-convective scaling property previously discussed.

When examining the response and the correlation functions obtained at different values of $Re_\lambda$, one first observes that, 
in agreement with the very good approximation provided by the analytical viscous Gaussian-convective formulae of 
Eq. (\ref{eq:resp-vgc}) to the response function, the latter is well described as an universal function of the adimensional 
variable $\tau \kappa u_0$. The same observation does not hold true for the normalized correlation function 
$\Qav(\kappa,\tau) / \Qav(\kappa,0)$ which once plotted in convective scaling shows a residual dependence on $Re_\lambda$. 
In particular when increasing $Re_\lambda$, the correlation function moves towards the response function, and this implies that 
the approximation involved by the classical FDR, Eq. (\ref{eq:scalfdr}), is gradually improving. For completeness the two functions 
$\Gav(\kappa,\tau)$ and $\Qav(\kappa,\tau)/\Qav(\kappa,0)$ are also plotted in terms of Kolmogorov viscous units, as shown in 
Fig. \ref{fig:NS_FDR_kd} (bottom). When this scaling is employed, neither the response nor the correlation show a collapse. 

%%%%%%%%%%%%%%%%%%%%%%
\section{Conclusions}
\label{sec:concl}

The mean linear response function of homogeneous isotropic turbulence to an impulsive body force has been measured through a number of numerical experiments carried out with DNS, at low and moderate values of the Reynolds number $Re_\lambda$. The measurement method leverages a white-noise forcing to probe the flow within the linearity constraint while mantaining the computational effort at reasonable levels. The method employed for measuring the response of the full turbulent flow has been thouroughly validated by computing the response function for purely viscous dynamics: the very same procedure yields this simpler response, for which an exact analytical expression is available to compare with. Based on this test case, the proper convergence with respect to the parameters of the time discretization has been verified. The methodology proposed here for measuring the response function has then proved effective in the quantitative description of the whole time decay of the response within the universal equilibrium range of scales. Our results have been verified both in terms of linearity of the response with respect to the amplitude of the forcing and adequateness of the time averaging. The same direct numerical simulations have been additionally employed for determining the turbulence correlation function. Its examination within the same range of scales has allowed us to preliminarly address the approximations involved by the classical fluctuation-dissipation relation when applied to turbulence dynamics. 

The analysis of the response function in the universal dissipative subrange confirms the theoretical prediction of energy-convective scaling for both the response and the normalized correlation functions, and, as shown by Figs. \ref{fig:NS_resp_1D_conv_scaled} and \ref{fig:NS_Q1D_conv_scaled}, establishes such scaling as the dominant one, at least in the rather limited range of $Re_\lambda$ considered here. A somewhat surprising result is that the analytical solution provided by Kraichnan in Ref. \cite{kraichnan-1964-b} to the problem of idealized convection turns out to be an extremely good approximation of the measured response function, with small deviations limited to the tail region, as shown in Fig. \ref{fig:NS_resp_1D_keta}. 

When comparing the normalized correlation function and the response function, a longer decorrelation time is observed for the former, as suggested by the theoretical arguments put forward in Ref. \cite{leslie-1973}. Both $\Gav(\kappa,\tau)$ and $\Qav(\kappa,\tau)/\Qav(\kappa,0)$ obey the same convective temporal scaling within the dissipation range, hence the two time scales $\tau_{\Gav}(\kappa)$ and $\tau_{\Qav}(\kappa)$ are in an approximately constant ratio. Obviously the Gaussian form of the FDR, Eq. (\ref{eq:scalfdr}), is not exactly satisfied, as witnessed from the departure between $\Gav(\kappa,\tau)$ and $\Qav(\kappa,\tau)/\Qav(\kappa,0)$ in Fig. \ref{fig:NS_FDR_kd}. Nevertheless, Eq. (\ref{eq:scalfdr}) remains a good approximation in terms of characteristic time scales, even at the moderate value of $Re_\lambda$ considered here and in the dissipation subrange, where less agreement would be expected in comparison to the inertial subrange, which is the proper context in which the FDR should be considered \cite{rosenblatt-vanatta-1972}. Moreover, the FDR approximation in the present range of scales appears to be increasingly better supported when the value of $Re_\lambda$ is increased, as shown in Fig. \ref{fig:NS_FDR_kd}. This last conclusion is in partial agreement with a previous study by Biferale {\em et al.} \cite{biferale-etal-2002}, who examined the response function within the inertial range of scales as extracted from the shell model. In that work the concept of halving-time statistics was introduced to better characterize the time properties of the response function for lower shells, where the proper $\tau$-convergence of the response cannot be easily achieved. The ratio between characteristics times is still constant in the inertial range, but $\tau_{\Qav}(\kappa)$ and $\tau_{\Gav}(\kappa)$ show Kolmogorov inertial time scaling. Both Kraichnan's arguments on random convections effects as well as the more recent and related discussion on the validity of the FDR in the context of turbulence, Ref. \cite{kraichnan-2000}, are strongly supported by present results. However Kraichnan indicates that the dominance of energy-advection effects on both the Eulerian response and the correlation functions are expected to extend to the dissipation range only at high values of Reynolds number, while this has been found in the present work to happen already at low or moderate values of $Re_\lambda$ addressed here. One possible explanation might be provided by considering the energy-convection effects as a feature of turbulence that remains limited to the dissipative subrange of scales, so that the presence of significant scale separation from the energy scales would let the random convection picture to hold: however the same could not be true for inertial scales at higher $Re_\lambda$.

A more thorough description of the response function and of its relevant time scales, together with a precise assessment of the approximations involved by the classical FDR, obviously call for an extension of the present study towards much higher values of $Re_\lambda$, so that a well-defined inertial range can develop. When such data will be available, the question about a possible asymptotic vanishing of the convective scaling in favor of a true Kolmogorov scaling could be properly answered, thus enlightening the framework of Eulerian closure theories. To this purpose, an analysis using halving-time statistics can be exploited to accurately characterize the properties of the response function in time over a wide range of scales. If Kolmogorov scaling will indeed be recovered at higher $Re_\lambda$, then the local relaxation processes of the turbulent response would be captured, opening a new scenario in the understanding of turbulence physics and modeling. 

%%%%%%%%%%%%%%%%%%%%%%%%%%
\begin{acknowledgments}
The authors would like to thank Dr. F. Martinelli for suggesting the Stokes test case described in \S\ref{sec:stokes} and for the helpful discussions. We gratefully acknowledge the use of the computing system located at the University of Salerno and the discussions with Prof.  P. Luchini.
\end{acknowledgments}

%\bibliographystyle{plain} 
%\bibliography{../../mq}

\end{document}